\documentclass[aps,prl,twocolumn,showpacs,groupedaddress]{revtex4}
\usepackage{graphicx}

\begin{document}

\title{Hidden Defect Pairs: Objects Invisible  in Low-Energy
Electron Scattering}

\author{A.\,A.\,Gorbatsevich}
\email{aag@qdn.miee.ru}
\affiliation{Moscow State Institute of Electronic Technology
(Technical University)\\
Zelenograd, Moscow, 124498, Russia}

\date{\today}

\begin{abstract}
Objects composed of lattice defects exist within a one-dimensional tight-binding model whose electron reflection coefficient in the low-energy case is equal to zero. Localized states are absent as well. The effective mass concept explains this not as some kind of reflectionless potential but as homogeneous medium, in which effective object size collapses. Without making effective mass approximations a new type of resonance is observed, in which the reflection coefficient becomes zero at a certain energy.

\end{abstract}


\maketitle

Electron scattering is a useful tool for studying atomic as well as condensed matter objects. The internal structure of these objects can manifest itself through various kinds of resonances. The long-wave-length case of the effective mass model (EMM) allows treating condensed matter as continuous medium. In this case all known resonances of elastic electron scattering  have direct optical (or microwave) analogues originating from the interference of waves scattered in extended reflecting systems: resonant tunneling and Fabri-Perro resonances, virtual-state resonances in over-barrier electron transmission in semiconductor heterostructures (Ramsauer-Townsend-like resonances) and resonances in light transmission through thin refracting plates, Fano resonances both in electron and microwave multimode waveguides.

In this letter we demonstrate the existence of a new type of electron transmission 
resonance. We show that extended microscopic objects can exist in a one-dimensional
lattice model which possess anomalous transparency in the case of electron scattering - they are invisible in the low-energy case. These objects are pairs of defects (hidden defect pairs - HDP) with antisymmetric defect potentials located at unequivalent sites at certain distances and in a special pattern. The transparency coefficient of an HDP equals unity without any phase factors and bound states are absent. This fact sets the described objects apart from the well-known reflectionless potentials. In the continuous case described by EMM the effective size of an object can essentially differ from its real microscopic size. Anomalous transparency of an HDP corresponds to the collapse of effective defect pair size in the continuous case. The continuous representation of a separate defect is known to be a delta-function potential. However an HDP in this case may appear as  a structureless homogeneous medium - as if the two defects cancel each other. Without the use of effective mass approximation microscopic objects also can be characterised by an effective size which in this case becomes energy dependent. At a certain energy this size becomes zero and a resonance of a new type takes place manifesting itself through the absence of electron reflection.

Consider a one-dimensional tight-binding model with both bond and
on-site alternation.
Different atoms in the unit cell are denoted by index $j$
($j=1,2$). Let two defects
(impurities) be located at sites $M_1 $ and $M_2$, occupied in a
host lattice by
atoms 1 and 2. The Hamiltonian of the model is:
\begin{eqnarray}
\hat H =& &\sum\limits_{j=1,2\;n_1,\; n_2}
[\varepsilon _{j\;}C_{j\:n_j}^\dag  C_{j\:n_j}  -
C_{1\:n_1}^\dag (t_ + C_{2\:n_1 + 1}+ \nonumber\\
&&+ t_ - C_{2\:n_1 - 1} )+
\varepsilon _{j}^* C_{M_j }^\dag  C_{Mj}+ h.c.]. \label{h1}
\end{eqnarray}
Here $C_{j\:n_j}^\dag\;(C_{j\:n_j})$ is the electron creation (annihilation) operator at site $n_j$, $n_1=2m,\;n_2=2m+1$ ($m$ - integer),
$\varepsilon _{1,2} $ are on-site energies of bulk material, while
$\varepsilon _{1,2}^* $ are on-site defect energies and
$t_{\pm}=t_1 \pm t_2$  are alternating hopping integrals. 
Hopping integrals for defect atoms are taken to be the
same as in the bulk. The model (\ref{h1}) describes $\Delta _{3(4)}
$ hole states originating
from atomic p-orbitals at the Brillouin zone (BZ) center in
semiconductors with
zinc-blend structure \cite{Cardona,comment1} as well as Peierls
insulators with both bond
and site alternation (asymmetric Peierls model). A simplified
version of this
model with $\varepsilon_1 =\varepsilon_2$ (symmetric Peierls model)
was successfully used to describe electronic states in
polyacetylene (PA)
\cite{Heeger}. The energy spectrum of the model (\ref{h1}) without
defects possesses two
(lower (L) and upper (U)) branches:
$E_{U,L} (k) = \varepsilon _{0\;}  \pm \sqrt {\delta^2  +
4t_1^2 \cos ^2 (ka) + 4t_2^2 \sin ^2 (ka)}
=\varepsilon _{0\;} \pm E(k)\ $. Here $\varepsilon _{0} =
 \frac{1}{2}(\varepsilon _{1\;}  + \varepsilon _{2\;} ),\;\delta
_{0}
 = \frac{1}{2}(\varepsilon _{1\;}  - \varepsilon _{2\;} )$ and
$\:a\:$-
is the interatomic distance, which we take to be
equal for adjacent bonds. For $t_2  > t_1$ the energy band extrema are
at the BZ center.
The wave functions for the upper band are
$c_{1,2k\;U} = u_k, -v_k e^{ - i\varsigma _{1,2}(k)}$, where the phase
factors
$\varsigma _{1,2}(k)$ are related to each other by the expression
\begin{equation}
\varsigma _{2\;} (k) - \varsigma _{1\;} (k) = \vartheta (k), \quad
\vartheta (k) = \arctan \left( \theta \tan(ka) \right),\label{f}
\end{equation}
with $\theta = t_2 /t_1$. Coefficients $u$ and $v$ have the
standard form of canonical transformation coefficients:
$u_k,v_k  =[(1/2)(1 \pm \delta /E(k))]^{\frac{1}{2}}$
\cite{comment2}.
For the lower band the solution is obtained by reversal: $u
\to -v, v \to u$.

Consider a scattering solution of the Hamiltonian (\ref{h1}), in the
form:
$c_{1,2\:n} = c_{1,2k}e^{ikna}  + r\,c_{1,2 - k}e^{ - ikna} ,
\;n < M_1;  \quad
c_{1,2\:m}  = tc_{1,2k} e^{ikma} \;,\;m > M_2$ .
After some algebraic manipulations we obtain the following expressions for the reflection ($r$) and transmission ($t$) coefficients in the upper band:

\begin{widetext}
\begin{equation}
 r =  - e^{2ikM_1 - 2i\varsigma _a } \frac{(\Delta _1 \Delta _2
u_k v_k+i\Delta_{k\:-} U_k)\sin [L_{eff}(k)k] - \Delta_{k\:+} U_k
\cos [L_{eff}(k)k]}
 {\Delta _1 \Delta _2 u_k v_k \sin [L_{eff}(k)k]-(\Delta_{k\:+} +
2iu_k v_k U_k)U_ke^{-iL_{eff}(k)k}}, \label{r1}
\end{equation}
\begin{equation}
 t = -e^{-iL_{eff}(k)k} \frac{{2iu_k v_k U_k^2 }}{\Delta _1 \Delta
_2 u_k v_k \sin [L_{eff}(k)k]-
 (\Delta_{k\:+}+2iu_k v_k U_k)U_ke^{-iL_{eff}(k)k}},  \label{t1}
\end{equation}
\end{widetext}
where
\begin{eqnarray*}
&\Delta _{1,2}  = \varepsilon _{1,2}^*  - \varepsilon _{1,2},\quad
\Delta_{k\:\pm}=\Delta _1 v_k^2  \pm \Delta _2 u_k^2 ,\\
&U_k = t_2\sin[\vartheta (k)]\cos(ka)-t_1\cos[\vartheta (k)]\sin(ka).
\end{eqnarray*}
$\Delta_j >0\:(<0)$ corresponds to donor- (acceptor-) like
impurity. In (\ref{r1}) the effective size of the defect pair is introduced.
\begin{equation}
L_{eff}(k)=L_0-\frac{1}{k}\vartheta (k),
\label{l1}
\end{equation}
where $L_{0}=(M_2  - M_1)a $. Under the condition
\begin{equation}
\Delta _1 v_k^2  =  - \Delta _2 u_k^2 
\label{d1}
\end{equation}
$\Delta_{k\:+}$ and the last term in the numerator of (\ref{r1}) both become 
zero and the reflection coefficient becomes proportional to
$\sin [L_{eff}(k)k]$. At $L_{eff}(k)=0$ \emph{resonance} takes
place: $r=0$ and $t=1$. In the symmetric Peierls model 
($\delta = 0$) the condition (\ref{d1}) is energy
independent and becomes $\Delta_1 =-\Delta_2$. In the asymmetric Peierls model the $u_k$ and
$v_k$ factors in (\ref{d1}) depend upon energy and the resonance is
observed only if the conditions (8) and $L_{eff}=0$ are satisfied
at one and the same energy.

In the case
\begin{equation}
ka <  < 1,\quad \vartheta (k)\approx \theta ka <  < 1,
\label{m1}
\end{equation}
which corresponds to the values of $k$ within EMM $L_{eff}(k)=L_{eff}=L_0 -
a\theta$ then the function $\sin[L_{eff}(k)k]$ in (\ref{r1}) becomes 
$\sin[(L_0 - a\theta)k]$.
Hence at large enough value of $\theta $ (or at small enough value
of $L_0$) such
that $L_{eff}=0$ the reflection coefficient $r$
identically equals zero
at all values of $k$ within the framework of EMM. 

For $t_1>t_2$ (as is the case for PA) the energy band extrema are at the BZ
boundary.
The continuous (EMM) case for such a model corresponds to small wave
vectors measured from
the extremum points at $k=\pm \pi /a$. It can be shown that for
such wave-vectors formulas
(\ref{r1}), (\ref{t1}) are maintained with the following modifications:
hopping integrals $t_1$
and $t_2$ are interchanged (parameter $\theta = t_1/t_2$) and the
reflection coefficient
acquires insignificant phase factor. 

The dependence of the effective length $L_{eff}(k)$ on the wave-vector  is shown
in Fig. 1. The parameter $\theta$ is usually accepted to be $7.1$ in band-structure 
calculations for PA \cite{Heeger}. In group $IV$ elements and $A3B5$ compounds this 
parameter varies from $2$ to $3$ \cite{Cardona,Vogel}.

\begin{figure}
\centerline{
\includegraphics
[width=8cm,height=6cm]
{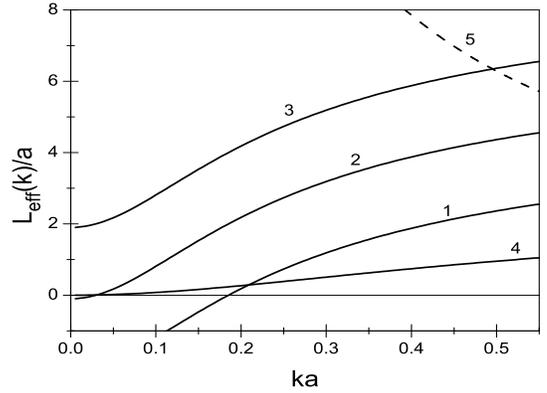}}
\caption{\label{fig1}
Dependence of effective interdefect distance
$L_{eff}(k)$ on wave-vector.  Curve $1 \;-\; \theta =7.1, L_0 =5;\;2\;-\; \theta
=7.1, L_0=7;\; 3\;-\;\theta =7.1, L_0=9;\; 4\;-\;\theta =3,L_0=3$.
Curve 5 - $kL_{eff}(k)=\pi $.}
\end{figure}

To understand the physical nature of the described
phenomenon consider continuous case where a point defect 
can be reproduced by a $\delta $-function potential.
For two $\delta $-functions located at points $x_1 $ and $x_2 $
it can be shown that in continuous EMM the reflection coefficient $r$
exactly matches the result (\ref{r1}) in the case (\ref{m1}) with $L_{eff}(k)$
being replaced by $x_2 - x_1$. For equal in magnitude and reverse in sign 
$\delta-$potentials $r$ exhibits usual virtual state resonance at $k(x_2-x_1) =\pi$ 
but it becomes identically zero if and only if $x_1=x_2$. Hence the 
analogous behavior of the reflection coefficient in
the discrete model (\ref{h1}) can be interpreted as if the effective size
of defect pair in the continuous case is determined not by real
physical defect separation $L_0$, but rather by the effective
length $L_{eff}(k)$ (\ref{l1}), which  can undergo collapse at certain 
values of the parameters.

One-dimensional objects with $r\equiv 0$ (reflectionless potentials) are known
in physics \cite {Landau, Mora}. However the transmission coefficient of a
reflectionless potential possesses a nonzero phase-factor: $t=e^{i\phi}$. 
The phase $\phi$ is determined by the location
of scattering amplitude poles in the complex wave-vector plane which
are related to energies of bound states. The phase of the transmission coefficient 
as well as the bound states can be used to detect the
existence of reflectionless potential. The potentials with zero effective size 
described in the present paper are much more hidden objects. The transmission 
coefficient $t$ (\ref{t1}) at the condition
$L_{eff}(k)=0$ (at $\Delta_+=0$) equals unity without any phase
factor. Bound states don't
exist either: it can be shown that dispersion relation for bound
state energy $r^{-1}(k=i\kappa)=0$ has no nontrivial solutions at
$L_{eff}(k)=0$. Hence HDP are completely invisible in the long-wave-length case. In
this case they appear as structureless homogeneous medium and in this sense
they have no optical analogue. On the other hand analogous objects can
be constructed in photonic crystals with dielectric permitivity periodically varying in space. 
Taking into account deviations from continuous case (nonparabolicities)
HDP manifest themselves as broad resonances at anomalously low energies.

Consider two defects located at equivalent lattice sites $M_1$ and $M_2$
(e.g. both occupied in a host lattice by atoms of sort 1) with energies
$\epsilon_1^*=\epsilon_1 +\Delta_1$ and $\epsilon_2^*=\epsilon_1
+\tilde \Delta_1$ respectively. It can be shown that the reflection coefficient 
$\tilde r$ in this case can be obtained from the expression (\ref{r1}) with
the following substitutions:
\begin{equation}
\tilde{r}(k) = r(k,\Delta_2 u_k \to \tilde{\Delta}_1 v_k,U_k v_k
\to U_k u_k,
L_{eff}(k)\to L_0) .
\label{r2}
\end{equation}
The main difference between (\ref{r2}) and (\ref{r1}) is that in (\ref{r2}) 
the effective length $L_{eff}(k)$ is replaced by the physical microscopic 
length $L_0=(M_2  - M_1)a$.
Hence any anomalies related to the possible collapse of a defect pair in the continuous 
case are absent.

The square modulus of the reflection coefficient $|r|^2$ in the symmetric Peirls model with
antisymmetric defect potential  as a function of wave-vector is shown 
in Fig.2. The parameters of curve 4 satisfy the condition $L_0=\theta$ 
of exact defect pair collapse in macroscopic case $ka<<1$. In this case $r(k=0)=0$. All the
other curves become unity at $k\to 0$. From Fig.2 one can see that even in the model where
$L_{eff}(k)$ doesn't become zero (curve 3) the decrease of the effective
defect pair size results in the formation of a pronounced minimum in the $r(k)$ curve at small
energies. The parameters for curve 5 are the same as for curve 2 except for location of
right atom which is shifted half a lattice period to the right. 
The resonance observed in curve 3 at $ka=0.495$ is a virtual state resonance
corresponding to the intersection of curves 3 and 5 in Fig.1.
The resonance in curve 5 is also a virtual state resonance at $k=\pi /L_0$ ($L_0=8a$).

\begin{figure}
\centerline{
\includegraphics
[width=8cm,height=6cm]
{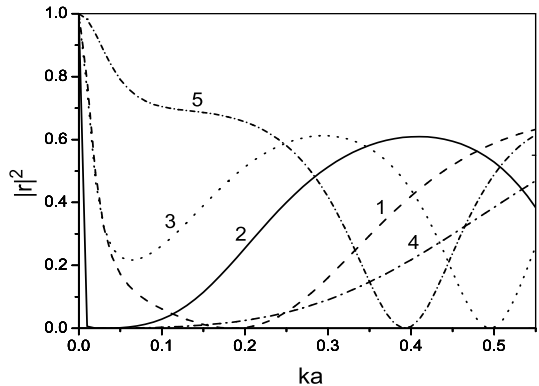}}
\caption{\label{fig2}
Dependence of reflection coefficient on wave-vector in SPM with
$\delta=0,\:t_1=3.2eV\:(t_1>t_2)$. For curves 1-4 other parameters
and labeling are the same as in Fig.1. Curve 5 corresponds
to pair of defects at equivalent lattice sites (\ref{r2}), $\theta
= 7.1,\:L_0=8a$.}
\end{figure}

The results obtained above mean that the location of the defect in continuous model 
can essentially differ from its physical microscopic position. The description of 
a point defect in the continuous regime is a particular case of an interface description.
An interface can be characterized in continuous EMM  by means of
a transfer matrix $\hat T$ which connects wave functions and their 
derivatives on both sides of the interface located at $x_0$ \cite{Ando,Tokatly}:
\begin{equation}
\left( {\begin{array}{*{20}c}
   \varphi   \\
   {\nabla \varphi }  \\
\end{array}} \right)_ +   = \hat T\left( {\begin{array}{*{20}c}
   \varphi   \\
   {\nabla \varphi }  \\
\end{array}} \right)_ -   = \left( {\begin{array}{*{20}c}
   {T_{11} } & {T_{12} }  \\
   {T_{21} } & {T_{22} }  \\
\end{array}} \right)\left( {\begin{array}{*{20}c}
   \varphi   \\
   {\nabla \varphi }  \\
\end{array}} \right)_ -
\label{bc}
\end{equation}

The parameters of the transfer matrix $\hat T$ can be calculated
using an extrapolation of the site amplitudes at the interface location 
\cite{Ando,Ivchenko}. More rigorously the off-diagonal elements 
of $\hat T$ can be calculated in a pseudopotential model \cite{Foreman}. 
However to get information about the precise location of the interface you 
should use another approach. 
Let's express the scattering data ($r$ and $t$) as a power series in wave-vector
for both an EMM with boundary condition (\ref{bc}) and a microscopic model. Within the EMM 
take only terms up to the first order of magnitude. Equate the coefficients at the 
terms of the same order in micro- and macroscopic case and obtain the transfer matrix 
parameters from the scattering data for microscopic model. You can also deduce in 
this way the interface location $x_0$ which enters the equations for scattering data 
in linear combination with the wave-vector.
In particular $T_{21}$ can be obtained from the expression for the
transmission coefficient at zero wave-vector of the incident wave:
$T_{21}=-2i{kt^{-1}}_{k\to 0}$.
It can be shown that for a heterointerface of two different
materials and nonzero (and
not small) $T_{21}$ in (\ref{bc}) the number of unknown parameters
is greater than the number
of equations resulting from the extraction procedure. Hence the interface
location $x_0$ can't be
determined. The situation is different for a solitary defect. In
this case $T_{11}=T_{22}=1$
and the number of unknown parameters is reduced. Let the defect in
microscopic model be located at site $M_1$, which we choose as a
coordinate origin: $x_0=M_1a+\Delta x_0$.
Decomposition of the reflection coefficient in EMM (\ref{bc}) for the plane
wave $e^{ik(x - \delta )} $ ($\delta $ - phase factor) incident from
the left in terms of the transfer matrix parameters then takes the
form:
\begin{equation}
r \approx   - e^{ikM_1a} [1 + 2ik(\frac{1}{{T_{ 21} }} + \Delta
x_0 - \delta ) + ...].
\label{r3}
\end{equation}
The reflection coefficient $r$ in the microscopic model \ref{h1} for the
defect replacing atom 1 of
host material is:
\begin{equation}
r =  - e^{2i(kMa - \varsigma _1 )} \Delta _1 v_k/(\Delta _1 v_k  -
2iu_kU_k),
\label{r4}
\end{equation}
where the parameters $\Delta_1,\:u_k$ and $U_k$ are the same as in
(\ref{r1}).
From the expression for the transmission coefficient one gets
$T_{21}=(v_0\Delta _1 t_1)/(u_0 (t_2 ^2-t_1 ^2)a)$ (here we assume
$t_2>t_1$).
Comparison of (\ref{r3}) with the decomposition of (\ref{r4}) in
powers of $ka$ provides for
the location of the defect: $x_{01}^{(1)} -\delta =M_1 a-
\varsigma_1.$
The same procedure repeated for the defect replacing the type 2 atom
located at site $M_2$ gives:
$ x_{02}^{(2)} -\delta =M_2 a- \varsigma_2$. Subtracting
$x_{02}^{(2)}$ from
$x_{01}^{(1)}$ eliminates the unspecified parameter $\delta$ and
obtains for the
effective length $L_{eff}=x_{02}^{(2)}-x_{02}^{(1)}=(M_2-M_1-
\theta )a$ - the
same expression as (\ref{l1}) in the case (\ref{m1}). If both atoms
are in equivalent
positions then their effective separation in the continuous case exactly coincides with their
real physical separation in the microscopic model $x_{02}^{(1)}-x_{01}^{(1)}=(M_2-M_1)a$.
This fact explains the radical difference between the reflection
coefficients of defect pairs (\ref{r1}) and (\ref{r2}) when their sizes differ only by
half a lattice period.
The unknown constant - phase factor $\delta$  can be determined by
applying an extraction procedure to the
heterointerface. In this case additional information can be
obtained from the decomposition of $r'$ in powers of wave-vector ($r'$ is the reflection 
coefficient for the wave incident from the right). As a result you obtain $\delta
=(\varsigma_1 + \varsigma_2 )/2$ and $ x_{0}^{(1,2)}=Ma \pm \theta /2.$
Therefore defects located at equivalent (nonequivalent) sites
shift in macroscopic picture in the same (opposite) direction(s).
$\Delta_3$ and $\Delta_4$ states in group $IV$ and $A3B5$ semiconductors
differ by the sign of $\theta$. Hence the collapse of a defect pair for the
$\Delta_3$ states will be accompanied by a doubling of the effective defect 
pair size for the $\Delta_4$ states and vice versa.

The physical nature of the described effect lies in the space-inversion 
asymmetry of the model (\ref{h1}). 
This microscopic asymmetry causes asymmetric wave function
distortions (with respect to the distribution of wave-function in the 
virtual reference symmetric model).
Within the framework of continuous EMM approach condensed matter
objects are structureless entities and their description is the same for both
symmetric and asymmetric microscopic models. Hence the only way to describe these
microscopic asymmetric distortions in the continuous case is to shift the location of 
the defect in respect to its position in the reference symmetric case.

To conclude, the extended microscopic object invisible in low-energy electron 
scattering - HDP - was constructed. Favorable conditions for the formation and
observation of an HDP can be provided by electrical annealing:
prolonged current transmission.
The association of separate defects into an HDP results in a local
decrease of
resistivity. In regions with smaller HDP concentration and higher
resistivity
Joule heat would enhance diffusion which helps HDP formation with
subsequent
reduction of resistivity and diffusion in this region. Finally
the whole specimen can transform into HDP enriched phase.
These pairs can be distributed chaotically and form a disordered conductor
without localization (a hidden defect structure).
Anomalous macroscopic behavior of scattering data results from
the underlying spatial asymmetry. This scenario seems to be quite
universal  and proposes a variety of "invisible" objects to exist in condensed matter. 
Another objects of this type are a quantum well in the model (\ref{h1}) with 
foreign atoms at the heterojunctions and a defect pair in the generalized Kronig-Penney 
model without center of inversion, which will be described in a separate paper. 

\begin{acknowledgments}
The author acknowledges the support of Russian Foundation for Basic
Research and Russian Ministry of Science and Education and helpful discussions 
with I.V.Tokatly.
\end{acknowledgments}


\begin{thebibliography}{99}
\bibitem {comment2}
Amplitudes $c_{k,\;k+\pi /a}$ usually used in Peierls model for PA
description  are: $c_{k,\;k+\pi
/a}=\frac{1}{\sqrt{2}}(c_{1\:k}\pm c_{2\:k}).$
\end{thebibliography}
\end{document}